\documentclass[sigconf]{acmart}
\acmConference[EASE 2026]{The 30th International Conference on Evaluation and Assessment in Software Engineering}{9–12 June, 2026}{Glasgow, Scotland, United Kingdom}

\AtBeginDocument{%
  }

\usepackage{pgfplots}
\pgfplotsset{compat=1.18}
\usepackage{enumitem}
\usepackage{hyperref}
\usepackage{multirow}
\usepackage{pifont}
\usepackage[htt]{hyphenat}
\usepackage{graphicx} 
\usepackage{listings}
\usepackage{xcolor}
\usepackage{tikz}
\usetikzlibrary{shapes.geometric, arrows.meta, positioning, fit, backgrounds}

\definecolor{keywordcolor}{RGB}{0, 112, 192}
\definecolor{stringcolor}{RGB}{163, 21, 21}
\definecolor{commentcolor}{RGB}{0, 128, 0}
\definecolor{backgroundcolor}{RGB}{248, 248, 248}

\lstdefinestyle{pythonstyle}{
    language=Python,
    basicstyle=\ttfamily\footnotesize,
    keywordstyle=\color{keywordcolor}\bfseries,
    stringstyle=\color{stringcolor},
    commentstyle=\color{commentcolor}\itshape,
    backgroundcolor=\color{backgroundcolor},
    frame=single,
    numbers=left,
    numberstyle=\tiny\color{gray},
    breaklines=true,
    showstringspaces=false,
    tabsize=4,
    morekeywords={self, True, False, None, pass}
}

\copyrightyear{2026}
\acmYear{2026}
\setcopyright{cc}
\setcctype{by}
\acmConference[EASE '26]{International Conference on Evaluation and Assessment in Software Engineering}{June 09--12, 2026}{Glasgow, United Kingdom}
\acmBooktitle{International Conference on Evaluation and Assessment in Software Engineering (EASE '26), June 09--12, 2026, Glasgow, United Kingdom}
\acmDOI{10.1145/3816483.3816547}
\acmISBN{979-8-4007-2348-3/2026/06}

\begin{document}
\title{\textsc{OpenClassGen}: A Large-Scale Corpus of Real-World Python Classes for LLM Research}

\author{Musfiqur Rahman}
\affiliation{%
    \institution{Concordia University}
    \city{Montr\'eal}
    \country{Canada}
}
\email{musfiqur.rahman@mail.concordia.ca}

\author{SayedHassan Khatoonabadi}
\affiliation{%
    \institution{Concordia University}
    \city{Montr\'eal}
    \country{Canada}
}
\email{sayedhassan.khatoonabadi@concordia.ca}

\author{Emad Shihab}
\affiliation{%
    \institution{Concordia University}
    \city{Montr\'eal}
    \country{Canada}
}
\email{emad.shihab@concordia.ca}

\renewcommand{\shortauthors}{Rahman et al.}

\begin{abstract}
Existing class-level code generation datasets are either synthetic (ClassEval: 100 classes) or insufficient in scale for modern training needs (RealClassEval: 400 classes), hindering robust evaluation and empirical analysis. We present \textbf{OpenClassGen}, a large-scale corpus of 324,843 Python classes extracted from 2,970 engineered open-source projects. Each entry pairs a human-written class with its corresponding skeleton, which comprises class and method signatures with associated docstrings, and is enriched with 27 static code metrics covering complexity, coupling, cohesion, and inheritance properties. Unlike prior benchmarks that require repository-level context resolution, OpenClassGen provides self-contained class skeletons that serve as complete generation specifications. We demonstrate the corpus's utility by evaluating three LLMs (GPT-o4-mini, Claude-4-Sonnet, Qwen-3-Coder) on a curated, executable subset of 300 classes, enriched with test suites achieving 58\% branch coverage. Results show strong semantic similarity (CodeBERTScore-F3: 0.89) but moderate functional correctness (pass rate: 0.33), with substantial variance across models. This variance, along with diverse class characteristics, confirms that OpenClassGen enables meaningful differentiation of LLM capabilities. The dataset supports diverse use cases, including fine-tuning, retrieval-augmented generation, difficulty modelling, and failure mode analysis. The complete dataset and curation scripts are publicly available at \url{https://zenodo.org/records/18409150}.
\end{abstract}



\maketitle


\section{Introduction}

Recent advancements in large language models (LLMs) have led to the development of various code generation benchmarks~\cite{athiwaratkun2022multi, austin2021program, chen2021evaluating, hendrycks2021measuring, iyer2018mapping}. However, these benchmarks predominantly focus on function-level code generation, evaluating LLMs on isolated, self-contained functions. At the other extreme, repository-level benchmarks such as RepoClassBench~\cite{deshpande2024class} and JavaBench~\cite{cao2024javabench} situate code generation within full project contexts, requiring complex agentic architectures to navigate codebases and retrieve relevant dependencies. This leaves an important middle ground underexplored: class-level code generation, where the structural complexity of object-oriented design is present, but the task remains tractable without heavyweight retrieval infrastructure.

Recent evidence suggests that this middle ground is particularly valuable. RepoClassBench reports up to 29\% pass rate in Python class generation with agentic architectures that have full repository access, while RealClassEval~\cite{rahman2025beyond} achieves comparable rates (25--34\%) without agents or repository-level context using only class skeletons. They also find that adding RAG-based retrieval yields up to a 7 percentage point increase in functional correctness. This suggests that well-designed class skeletons capture sufficient structural information for effective code generation. This offers a favourable trade-off between task complexity and infrastructure requirements. However, existing class-level datasets remain limited. ClassEval~\cite{du2024evaluating} provides 100 manually crafted Python classes, but its synthetic nature limits generalizability. For example, LLMs achieve 84--89\% correctness on ClassEval but only 25--34\% on real-world code~\cite{rahman2025beyond}. RealClassEval addresses realism by curating 400 classes from open-source projects, but this scale remains insufficient for training, fine-tuning, or large-scale empirical studies.

To address this gap, we present \textbf{OpenClassGen}: a large-scale corpus of 324,843 Python classes extracted from 2,970 engineered open-source projects. Each entry consists of a human-written class paired with its corresponding skeleton, comprising the class and method signatures with associated docstrings. We enrich the dataset with 27 static code metrics covering complexity, coupling, cohesion, and inheritance properties. Unlike prior datasets, OpenClassGen is designed not as a fixed benchmark but as a research infrastructure supporting multiple use cases:

\begin{itemize} [leftmargin=*]
    \item \textbf{Fine-tuning and training}: 324,843 (skeleton, implementation) pairs provide a supervised learning signal for class-level code generation.
    \item \textbf{Retrieval-augmented generation (RAG)}: Structural metadata enables similarity-based retrieval of relevant examples during generation.
    \item \textbf{Difficulty modelling}: Static metrics can be correlated with LLM performance to predict challenging class characteristics.
    \item \textbf{Empirical studies}: The scale supports statistically robust analysis of LLM failure modes, documentation effects, and architectural patterns.
\end{itemize}

This paper makes the following contributions:

\begin{enumerate} [leftmargin=*]
    \item \textbf{OpenClassGen}, the first large-scale corpus of real-world Python classes specifically curated for LLM-assisted class-level code generation research.
    \item A rigorous curation pipeline with engineering-based project filtering, AST-based skeleton extraction, and test class removal.
    \item An evaluation demonstrating dataset utility across three LLMs (GPT-o4-mini, Claude-4-Sonnet, Qwen-3-Coder) and four quality dimensions (syntactic, semantic, structural, functional).
    \item Public release of the dataset and replication scripts to support reproducibility and future research (See \nameref{sec:replication} section).
\end{enumerate}

The remainder of this paper is organized as follows. Section~\ref{sec:related} reviews related work. Section~\ref{sec:curation} details the curation process. Section~\ref{sec:characterization} characterizes the dataset. Section~\ref{sec:evaluation} presents our evaluation. Sections~\ref{sec:discussion} and~\ref{sec:limitations} discuss implications and limitations. Section~\ref{sec:conclusion} concludes the paper.

\section{Related Work}
\label{sec:related}

We review existing benchmarks for LLM-based code generation, focusing on the progression from function-level to class-level evaluation. We distinguish between \textit{class-level generation}, which concerns the granularity of the generated artifact (producing a complete class as the unit of output), and \textit{repository-level generation}, which concerns the context required for generation (resolving cross-file dependencies, imports, and project-specific configurations). While related, these dimensions are orthogonal: repository-level benchmarks may generate functions or methods that depend on external context, whereas class-level benchmarks focus on producing structurally complete classes regardless of external dependencies.

\subsection{Function-Level Benchmarks}

HumanEval~\cite{chen2021evaluating} is a widely used benchmark consisting of 164 hand-written Python programming problems with function signatures and test cases. It introduced Pass@k as an evaluation metric, measuring how often a model generates a correct solution within $k$ attempts. MBPP~\cite{austin2021program} extends function-level evaluation with 974 Python programs, each paired with natural language descriptions and test cases. Multi-HumanEval~\cite{athiwaratkun2022multi} builds on HumanEval by introducing multilingual code generation benchmarks covering over 10 programming languages. CoderEval~\cite{yu2024codereval} addresses context dependence by including non-standalone functions that rely on surrounding code, such as variables, functions, or classes within the same file or project. HumanEvo~\cite{zheng2024humanevo} addresses a different dimension of realism: temporal evolution. By rolling back repositories to the commit state before target functions were written, HumanEvo avoids future context leakage and reveals that evolution-ignored evaluation inflates LLM performance by 10--61\%.

While these benchmarks have advanced LLM evaluation for code generation, they remain limited to function-level tasks. Functions in isolation do not capture the complexity of real-world software, where code typically depends on class attributes, inheritance hierarchies, and project-specific dependencies.

\subsection{Class-Level and Repository-Level Benchmarks}

ClassEval~\cite{du2024evaluating} introduced the first benchmark specifically designed for class-level code generation, containing 100 manually crafted Python classes with test suites achieving 98.2\% branch coverage. The study revealed that LLMs perform significantly worse on class-level tasks compared to function-level benchmarks. For instance, GPT-4 achieved only 37\% holistic pass rate. The authors found that holistic generation (generating the full class at once) works better for advanced models, while incremental generation benefits weaker models. Yuen et al.~\cite{yuen2025prompting} extended this work, demonstrating that chain-of-thought prompting improves functional correctness by up to 25\%.

RealClassEval~\cite{rahman2025beyond} addressed the realism gap by curating 400 classes from GitHub repositories. The study found that LLMs achieve 84--89\% correctness on synthetic benchmarks but only 25--34\% on real-world code, with no significant generalization gap between seen and unseen repositories. Error analysis revealed that \texttt{AttributeError}, \texttt{TypeError}, and \texttt{AssertionError} dominate real-world failures. While RealClassEval advances ecological validity, its 400-sample scale remains insufficient for training, fine-tuning, or large-scale empirical analysis.

Adjacent to class-level generation, several benchmarks target repository-level code generation, where challenges involve resolving cross-file dependencies and navigating project context. JavaBench~\cite{cao2024javabench} addresses the language imbalance in existing benchmarks, noting that 95.8\% of code generation benchmarks involve Python. It comprises 4 Java projects originally designed as undergraduate course assignments, with 389 methods across 106 classes targeting object-oriented features such as encapsulation, inheritance, and polymorphism. RepoClassBench~\cite{deshpande2024class} extends evaluation to repository contexts across multiple languages (Java, Python, and C\#), where each class has cross-file dependencies and corresponding test cases. However, this benchmark requires sophisticated agentic architectures to retrieve relevant context and a dependency-resolution infrastructure, both of which add complexity and cost to evaluation.


\subsection{Positioning OpenClassGen}

Table~\ref{tab:comparison} compares OpenClassGen against existing datasets that target class-level or repository-level code generation. We include repository-level benchmarks (JavaBench, RepoClassBench) because they also address object-oriented code structures, though they differ in granularity and context requirements.

OpenClassGen fills a critical gap by providing two orders of magnitude more samples than prior class-level datasets while maintaining real-world origin. Unlike synthetic benchmarks (ClassEval), our classes are extracted from production repositories with varying documentation quality. Unlike existing real-world class-level datasets (RealClassEval), our scale supports fine-tuning, retrieval-augmented generation, and statistically robust empirical studies. Unlike repository-level benchmarks, OpenClassGen isolates the class-level synthesis problem: as noted in Section 1, RealClassEval achieves pass rates comparable to RepoClassBench's agentic approach using only class skeletons, suggesting that well-designed skeletons capture sufficient context for effective generation. Our corpus enables researchers to study this core challenge without the confounding factors of repository navigation. Additionally, we provide 27 static code metrics per class, enabling research on difficulty modelling and structural analysis not possible with existing datasets.

\begin{table}[t]
\centering
\caption{Comparison of class-level and repository-level code generation datasets.}
\label{tab:comparison}
\begin{tabular}{lrrc}
\toprule
\textbf{Dataset} & \textbf{Classes} & \textbf{Projects} & \textbf{Real-World} \\
\midrule
ClassEval~\cite{du2024evaluating} & 100 & --- & \ding{55} \\
RealClassEval~\cite{rahman2025beyond} & 400 & 12 & \ding{51} \\
JavaBench~\cite{cao2024javabench} & 106 & 4 & \ding{55} \\
RepoClassBench~\cite{deshpande2024class} & 130 & 26 & \ding{51} \\
\textbf{OpenClassGen} & 324,843 & 2,970 & \ding{51} \\
\bottomrule
\end{tabular}
\end{table}

\section{Dataset Curation}
\label{sec:curation}

This section describes the process followed to curate OpenClassGen. Our corpus targets Python due to its widespread adoption in both industry and open-source communities~\cite{madnightGitHubLanguage}, LLMs' strongest code generation performance on this language~\cite{chen2021evaluating}, and alignment with existing benchmarks, over 95\% of which target Python~\cite{cao2024javabench}. Figure~\ref{fig:pipeline} provides an overview of the curation pipeline.

\begin{figure*}[t]
\centering
\resizebox{\textwidth}{!}{
\begin{tikzpicture}[
    node distance=0.8cm and 1.2cm,
    box/.style={
        rectangle,
        draw=black,
        thick,
        fill=blue!10,
        minimum width=2.8cm,
        minimum height=1cm,
        align=center,
        font=\small
    },
    data/.style={
        rectangle,
        draw=black,
        thick,
        fill=green!15,
        minimum width=2.2cm,
        minimum height=0.8cm,
        align=center,
        font=\small
    },
    arrow/.style={
        ->,
        >=Stealth,
        thick
    },
    label/.style={
        font=\footnotesize,
        align=center
    }
]

\node[box] (step1) {Project\\Selection};
\node[data, above=0.5cm of step1] (input1) {GitHub\\Repositories};
\node[label, below=0.3cm of step1] (label1) {Engineered projects};

\node[box, right=1.5cm of step1] (step2) {Repository\\Analysis};
\node[label, below=0.3cm of step2] (label2) {Clone repos\\Extract 27 metrics};

\node[box, right=1.5cm of step2] (step3) {Class\\Extraction};
\node[label, below=0.3cm of step3] (label3) {AST parsing\\Skeleton generation};

\node[box, right=1.5cm of step3] (step4) {Post-Processing\\and Filtering};
\node[label, below=0.3cm of step4] (label4) {Remove AST failures\\Remove test classes};

\node[data, right=1.5cm of step4] (output) {OpenClassGen\\324,843 classes};

\draw[arrow] (input1) -- (step1);
\draw[arrow] (step1) -- node[above, font=\footnotesize] {2,970 repos} (step2);
\draw[arrow] (step2) -- node[above, font=\footnotesize] {434,518 classes} (step3);
\draw[arrow] (step3) -- (step4);
\draw[arrow] (step4) -- (output);


\end{tikzpicture}}
\caption{Overview of the OpenClassGen curation pipeline. Starting from 2,970 engineered GitHub repositories, we extract classes via AST parsing, generate skeletons, and apply filtering to produce 324,843 production Python classes.}
\label{fig:pipeline}
\end{figure*}

\subsection{Project Selection}

We build our dataset on existing filtered projects shared in~\cite{rahman2025beyond}. These projects were selected based on ``engineered project''~\cite{munaiah2017curating} criteria established by Xiao et al.~\cite{xiao2025self} to identify well-maintained, production-quality projects. Three categories of filters were applied:

\begin{enumerate} [leftmargin=*]
    \item \textbf{License filtering}: Repositories without licenses or with non-standard licenses were excluded, including those with Creative Commons licenses and the SIL Open Font License 1.1, which are not typically used for software projects.
    \item \textbf{Activity filtering}: Repositories without any releases, with fewer than two contributors, or marked as archived were removed.
    \item \textbf{Quality filtering}: The distribution of repository properties (pull requests, issues, and lines of code) was analyzed, and repositories in the first quartile for each metric were excluded. Additionally, repositories outside the 97\% confidence interval for code-to-comment ratio were excluded, as engineered software projects are typically documented with code comments.
\end{enumerate}

This process yielded 2,970 repositories spanning diverse domains, including web frameworks, machine learning libraries, and data processing utilities.

\subsection{Repository Analysis}

After identifying the projects, we cloned each repository, preserving the complete structure and dependencies of the codebases. This step ensured that all relevant files were available for analysis and maintained the integrity of the original repositories while enabling large-scale processing.

To characterize class complexity, we computed 27 static code metrics per class using Understand\texttrademark{} by SciTools~\cite{scitoolsUnderstandSoftware}. These metrics span four categories:

\begin{itemize} [leftmargin=*]
    \item \textbf{Size metrics}: Lines of code, comment lines, blank lines, number of methods.
    \item \textbf{Complexity metrics}: Cyclomatic complexity, essential complexity, maximum nesting depth.
    \item \textbf{Coupling metrics}: Fan-in, fan-out, number of dependencies.
    \item \textbf{Inheritance metrics}: Depth of inheritance tree, number of children, coupling between objects.
\end{itemize}

This analysis identified 434,518 Python classes across the 2,970 repositories.

\subsection{Class Extraction}

We parse all \texttt{.py} files using Python's \texttt{ast} module~\cite{pythonAbstractSyntax} to extract class skeletons, comprising class signatures, method signatures, and docstrings (when present), with method bodies replaced by \texttt{pass} statements.

Listings~\ref{lst:human} and~\ref{lst:skeleton} illustrate the extraction: the skeleton preserves structural information and documentation while removing implementations.

\begin{lstlisting}[style=pythonstyle, caption={Human-written class from nuagenetworks/monolithe~\cite{githubGitHubNuagenetworksmonolithe}.}, label={lst:human}]
class TaskManager(object):
    """Multi threading manager"""
    def __init__(self):
        """Initializes a TaskManager"""
        self.threads = list()

    def wait_until_exit(self):
        """Wait until all threads are finished."""
        [t.join() for t in self.threads]
        self.threads = list()

    def start_task(self, method, *args, **kwargs):
        """Start a task in a separate thread
        Args:
            method: the method to start
            args: Accept args/kwargs arguments
        """
        thread = threading.Thread(
            target=method, args=args, kwargs=kwargs)
        thread.is_daemon = False
        thread.start()
        self.threads.append(thread)
\end{lstlisting}

\begin{lstlisting}[style=pythonstyle, caption={Extracted class skeleton.}, label={lst:skeleton}]
class TaskManager(object):
    """Multi threading manager"""
    def __init__(self):
        """Initializes a TaskManager"""
        pass

    def wait_until_exit(self):
        """Wait until all threads are finished."""
        pass

    def start_task(self, method, *args, **kwargs):
        """Start a task in a separate thread
        Args:
            method: the method to start
            args: Accept args/kwargs arguments
        """
        pass
\end{lstlisting}

\subsection{Post-Processing and Filtering}

After extracting class skeletons, we performed post-processing to refine the dataset:

\begin{enumerate}[leftmargin=*]
\item \textbf{AST failure removal}: Classes that failed AST parsing were removed, reducing the dataset by 3,557 classes. Qualitative analysis revealed these typically involved Python 2 syntax (e.g., \texttt{print "..."} without parentheses), incompatible with Python 3's \texttt{ast} module.
\item \textbf{Test class removal}: Test classes were identified and removed using a multi-pattern heuristic on class names and file paths: class names starting or ending with ``Test'' or ``Case'', or located in paths containing ``test(s)'' or ``unit''. Test classes were excluded because they represent a fundamentally different generation task---they verify behavior rather than implement functionality, follow rigid framework-specific patterns (e.g., \texttt{unittest}, \texttt{pytest}), and depend on the classes they test. This step removed 106,118 classes. This ensures OpenClassGen focuses on production code generation, aligning with the scope of existing class-level benchmarks~\cite{du2024evaluating, rahman2025beyond}.
\end{enumerate}

\emph{The final dataset comprises 324,843 production Python classes.}

\subsection{Dataset Structure}

For each class, we record the fields described in Table~\ref{tab:dictionary}. The complete list of 27 static code metrics is available in the replication package.

\begin{table}[t]
\centering
\caption{Data dictionary for OpenClassGen.}
\label{tab:dictionary}
\begin{tabular}{p{0.28\columnwidth}p{0.65\columnwidth}}
\toprule
\textbf{Field} & \textbf{Description} \\
\midrule
\texttt{id} & Unique index for each class. \\
\texttt{repository\_name} & Name of the GitHub repository. \\
\texttt{file\_path} & Path to the file containing the class. \\
\texttt{class\_name} & Name of the class. \\
\texttt{human\_written\_class} & Complete class implementation with docstrings. \\
\texttt{class\_skeleton} & Extracted skeleton with method bodies replaced by \texttt{pass}. \\
\texttt{comment\_to\_code\_ratio} & Ratio of comment lines to code lines. \\
\texttt{total\_program\_units} & Number of classes and methods in the skeleton. \\
\texttt{total\_doc\_str} & Number of program units with docstrings. \\
\bottomrule
\end{tabular}
\end{table}

Table~\ref{tab:statistics} presents summary statistics for the curated dataset. Classes vary substantially in size and documentation coverage, reflecting the diversity of real-world software. The median class contains 24 lines of code with 1 method, while larger classes exceed 200 lines with complex method structures.

\begin{table}[t]
\centering
\caption{Summary statistics of OpenClassGen (324,843 classes).}
\label{tab:statistics}
\begin{tabular}{lrrrrr}
\toprule
\textbf{Property} & \textbf{Avg.} & \textbf{Std.} & \textbf{25\%} & \textbf{Med.} & \textbf{75\%} \\
\midrule
Method Count & 3.52 & 7.36 & 0.00 & 1.00 & 4.00 \\
Total Lines & 80.20 & 253.92 & 8.00 & 24.00 & 72.00 \\
Code Lines & 50.19 & 160.74 & 6.00 & 15.00 & 44.00 \\
Comment Lines & 20.86 & 102.85 & 0.00 & 3.00 & 15.00 \\
Blank Lines & 10.93 & 33.50 & 1.00 & 3.00 & 10.00 \\
Comment-to-Code Ratio & 0.53 & 1.71 & 0.00 & 0.22 & 0.63 \\
\bottomrule
\end{tabular}
\end{table}

\section{Dataset Characterization}
\label{sec:characterization}

To assess the quality and diversity of OpenClassGen, we analyze the dataset along three dimensions: documentation coverage, structural composition, and domain diversity.

\subsection{Documentation Coverage}

Documentation quality directly impacts LLM performance on code generation tasks, as docstrings provide natural language guidance for implementation. We categorize classes by documentation level:

\begin{itemize}[leftmargin=*]
    \item \textbf{Fully documented}: All program units (class and methods) have docstrings.
    \item \textbf{Partially documented}: Some but not all program units have docstrings.
    \item \textbf{Undocumented}: No program units have docstrings.
\end{itemize}

\begin{figure}[t]
\centering
\begin{tikzpicture}
\begin{axis}[
    ybar,
    bar width=0.8cm,
    width=0.85\columnwidth,
    height=5cm,
    ylabel={Percentage of Classes (\%)},
    symbolic x coords={Full docstr, Partial docstr, No docstr},
    xtick=data,
    xticklabel style={text width=2.2cm, align=center, font=\small},
    ymin=0,
    ymax=50,
    ytick={0, 10, 20, 30, 40, 50},
    nodes near coords,
    nodes near coords align={vertical},
    every node near coord/.append style={font=\small},
    enlarge x limits=0.25,
    grid=major,
    grid style={dashed, gray!30},
    axis lines*=left,
]
\addplot[fill=blue!60, draw=blue!80] coordinates {
    (Full docstr, 19.7)
    (Partial docstr, 44.4)
    (No docstr, 36.0)
};
\end{axis}
\end{tikzpicture}
\caption{Distribution of classes by documentation coverage in OpenClassGen.}
\label{fig:documentation}
\end{figure}

Figure~\ref{fig:documentation} shows the distribution across these categories. Approximately 20\% of classes are fully documented, 44\% are partially documented, and 36\% have no docstrings. This distribution reflects real-world software practices, where documentation coverage varies substantially across projects and developers~\cite{he2019understanding}. Unlike synthetic benchmarks such as ClassEval, which provide complete and well-structured docstrings for all classes, OpenClassGen captures the documentation variability that LLMs encounter in production environments.

The dataset includes fields \texttt{total\_program\_units} and \texttt{total\_doc\_str} for each class, enabling researchers to filter by documentation level. For instance, selecting classes where \texttt{total\_program\_units} equals \texttt{total\_doc\_str} yields 63,971 fully documented classes, which in and of itself is significantly larger than existing class-level benchmarks.

\subsection{Structural Composition}

Classes in OpenClassGen exhibit substantial structural diversity. Table~\ref{tab:statistics} shows that the median class contains 1 method and 24 lines of code, while the 75th percentile reaches 4 methods and 72 lines. This range spans simple data containers and utility classes to complex stateful managers with extensive method interactions.

We observe the following structural patterns:

\begin{itemize}[leftmargin=*]
    \item \textbf{Data classes}: Classes with few or no methods, primarily defining attributes. These constitute approximately 28\% of the dataset (classes with zero methods).
    \item \textbf{Utility classes}: Small classes with 1--3 focused methods providing specific functionality. These represent approximately 44\% of the dataset.
    \item \textbf{Manager classes}: Larger classes with 4+ methods managing state and coordinating complex operations. These comprise approximately 28\% of the dataset.
\end{itemize}

The 27 static metrics enable fine-grained filtering by structural properties. For example, researchers studying inheritance can filter by depth of the inheritance tree; those investigating complexity can stratify by cyclomatic complexity.

\subsection{Domain Diversity}

OpenClassGen spans 2,970 repositories covering diverse application domains. Based on qualitative analysis of the repositories, the dataset includes cloud and infrastructure tooling (14\%), scientific computing (8\%), data processing and ETL (6\%), machine learning and AI (6\%), DevOps and CI/CD (5\%), API clients and SDKs (5\%), networking libraries (4\%), web frameworks (2\%), and testing tools (2\%), with the remainder spanning domain-specific utilities across robotics, gaming, cryptography, and other specialized fields. This diversity ensures that models trained or evaluated on OpenClassGen encounter varied coding patterns, naming conventions, and architectural styles representative of the broader Python ecosystem.

\subsection{Comparison with Existing Datasets}

Table~\ref{tab:quality_comparison} compares OpenClassGen's characteristics with existing class-level datasets. We restrict this comparison to ClassEval and RealClassEval as they are the only datasets that share our granularity (class-level) and language (Python), enabling direct comparison of documentation coverage, structural metrics, and domain diversity. As shown in the table, OpenClassGen provides not only greater scale but also greater variability in documentation and structure, enabling more robust evaluation of LLM generalization.

\begin{table}[t]
\centering
\caption{Dataset characteristics comparison.}
\label{tab:quality_comparison}
\resizebox{\columnwidth}{!}{
\begin{tabular}{lccc}
\toprule
\textbf{Characteristic} & \textbf{ClassEval} & \textbf{RealClassEval} & \textbf{OpenClassGen} \\
\midrule
Classes & 100 & 400 & 324,843 \\
Documentation & Complete & Variable & Variable \\
Structural metrics & \ding{55} & \ding{55} & 27 metrics \\
Domain diversity & Limited & Moderate & High \\
\bottomrule
\end{tabular}}
\end{table}

\section{Evaluation}
\label{sec:evaluation}

To demonstrate the utility of OpenClassGen for class-level code generation research, we evaluate LLM-generated classes using extracted skeletons as structured prompts. This evaluation is explicitly illustrative: its purpose is to confirm that OpenClassGen supports meaningful multi-dimensional LLM assessment, not to provide definitive model rankings or characterize the state of the art. The three models selected are representative of current general-purpose and code-specialized LLMs, but are not claimed to be the strongest available baselines. We assess generation quality by comparing LLM outputs against their human-written counterparts across four complementary dimensions: syntactic quality (lexical overlap), semantic quality (meaning preservation), structural quality (AST-level alignment), and functional correctness (executable behaviour).

\subsection{Approach}

\paragraph{Sampling.} We randomly sample 300 classes from the dataset using stratified sampling to ensure balanced representation across three documentation levels: fully documented (all program units have docstrings), partially documented (some program units have docstrings), and undocumented (no program units have docstrings). This stratification enables evaluation of LLM performance under varying levels of natural language guidance. A sample of 300 classes (100 per documentation tier) exceeds the minimum required for a 90\% confidence level with a 5\% margin of error ($n = 271$), providing sufficient statistical resolution for this illustrative evaluation; a more comprehensive evaluation leveraging the full dataset's scale is an intended direction for future work.

\paragraph{Generation.} For each sampled class, we prompt three representative LLMs to generate the complete class implementation given its skeleton:

\begin{itemize} [leftmargin=*]
    \item \textbf{GPT-o4-mini}~\cite{openaiOpenAIPlatform}: A reasoning-focused model from OpenAI.
    \item \textbf{Claude-4-Sonnet}~\cite{anthropicIntroducingClaude}: A general-purpose model from Anthropic.
    \item \textbf{Qwen-3-Coder}~\cite{huggingfaceQwenQwen3Coder30BA3BInstructHugging}: A code-specialized open-weight model.
\end{itemize}

We use the following prompt:

\begin{quote}
\textit{You are an expert Python programmer who can correctly implement complete Python classes based on the provided class skeletons. Implement the following class:}

\texttt{[CLASS SKELETON]}
\end{quote}

This prompt design follows established practices in code generation evaluation~\cite{chen2021evaluating, du2024evaluating}. The ``expert'' persona encourages the model to produce professional-quality code rather than simplified or pedagogical implementations; prior work has shown that expert identity prompting improves LLM response quality across diverse tasks~\cite{xu2023expertprompting}. We intentionally avoid chain-of-thought instructions or few-shot examples to isolate the model's inherent class-level generation capability, ensuring results are comparable across models and reproducible by other researchers.

\subsection{Metrics}

We employ five metrics spanning four quality dimensions. We provide formal definitions below, following prior work in code generation evaluation~\cite{du2024evaluating, rahman2025beyond}.

\subsubsection{Syntactic Quality}

We measure lexical overlap using two complementary n-gram-based metrics. Following prior work~\cite{hindle2016naturalness, rahman2019natural}, we use $n \leq 3$ as higher-order n-grams provide diminishing returns for code representation.

\paragraph{BLEU} The BLEU (Bilingual Evaluation Understudy) score~\cite{papineni2002bleu} measures precision-oriented n-gram overlap between generated code $C$ and reference code $R$:

\begin{equation}
\text{BLEU} = BP \cdot \exp\left(\sum_{n=1}^{N} w_n \log p_n\right)
\end{equation}

\noindent where $p_n$ is the precision of n-grams up to length $N$, $w_n$ is the weight assigned to each n-gram (typically uniform), and $BP$ is a brevity penalty that adjusts for short generations:

\begin{equation}
BP = 
\begin{cases}
1 & \text{if } |C| > |R| \\
e^{(1 - |R|/|C|)} & \text{if } |C| \leq |R|
\end{cases}
\end{equation}

\noindent where $|C|$ and $|R|$ denote the lengths of the candidate and reference, respectively.

\paragraph{ROUGE-L} The ROUGE (Recall-Oriented Understudy for Gisting Evaluation) score~\cite{lin2004rouge} measures recall-oriented overlap. We use the ROUGE-L variant, which computes the longest common subsequence (LCS) between generated and reference code:

\begin{equation}
\text{ROUGE-L} = \frac{\text{LCS}(C, R)}{|R|}
\end{equation}

\noindent where $\text{LCS}(C, R)$ is the length of the longest common subsequence. Unlike n-gram metrics, ROUGE-L captures sequence-level structure without requiring contiguous matches.

Together, BLEU emphasizes precision (how much of the generated code is correct) while ROUGE-L emphasizes recall (how much of the reference code is covered).

\subsubsection{Semantic Quality}

\paragraph{CodeBERTScore} Lexical metrics fail to capture semantic equivalence when implementations differ syntactically but preserve meaning. CodeBERTScore~\cite{zhou2023codebertscore} addresses this by computing similarity in the embedding space of CodeBERT~\cite{feng2020codebert}, a transformer model pre-trained on code and natural language.

Given token embeddings $\mathbf{c}_i$ for generated code and $\mathbf{r}_j$ for reference code, CodeBERTScore computes precision $P$, recall $R$, and F-score:

\begin{equation}
P = \frac{1}{|C|} \sum_{c_i \in C} \max_{r_j \in R} \mathbf{c}_i^\top \mathbf{r}_j
\end{equation}

\begin{equation}
R = \frac{1}{|R|} \sum_{r_j \in R} \max_{c_i \in C} \mathbf{r}_j^\top \mathbf{c}_i
\end{equation}

\begin{equation}
F_\beta = (1 + \beta^2) \cdot \frac{P \cdot R}{\beta^2 \cdot P + R}
\end{equation}

We report $F_3$ (i.e., $\beta = 3$), which weights recall three times more than precision. This choice reflects that in code generation, capturing the full semantic content of the reference is more critical than avoiding extraneous tokens.

\subsubsection{Structural Quality}

\paragraph{Tree Similarity Edit Distance (TSED)} Lexical and embedding-based metrics operate on token sequences and may miss structural differences in code organization. TSED~\cite{song2024enhancing, song2024revisiting} measures structural similarity by comparing abstract syntax trees (ASTs).

Given ASTs $T_1$ and $T_2$ for generated and reference code, the tree edit distance $\Delta(T_1, T_2)$ is the minimum cost of edit operations (insertion, deletion, renaming) to transform $T_1$ into $T_2$:

\begin{equation}
\Delta(T_1, T_2) = \min_{\text{ops}} \sum_{i=1}^{n} w(op_i)
\end{equation}

\noindent where $ops$ is a sequence of edit operations and $w(op_i)$ is the cost of each operation. We compute this using the APTED algorithm~\cite{pawlik2016tree}, which optimizes over all possible path strategies.

To ensure interpretability, we normalize by the size of the larger AST:

\begin{equation}
\text{TSED} = 1 - \frac{\Delta(T_1, T_2)}{\max(|T_1|, |T_2|)}
\end{equation}

\noindent where $|T|$ denotes the number of nodes in tree $T$. TSED ranges from 0 (completely different structures) to 1 (identical ASTs).

\subsubsection{Functional Correctness}

\paragraph{Pass Rate} The above metrics assess similarity to reference implementations but do not verify executable correctness. We compute the pass rate on automatically generated unit tests:

\begin{equation}
\text{Pass Rate} = \frac{\text{Number of passing tests}}{\text{Total number of tests}}
\end{equation}

Tests are generated using Pynguin~\cite{lukasczyk2022pynguin}, a search-based test generation~\cite{panichella2018large} framework for Python that applies genetic algorithms to maximize code coverage~\cite{campos2018empirical}.

We use automated test generation rather than repository test suites for two reasons. First, executing repository tests at scale is impractical: it requires resolving project-specific dependencies~\cite{latendresse2022not}, handling environment mismatches, and managing unmaintained projects. These challenges do not scale to 324K classes across 2,970 repositories. Second, Pynguin generates tests against the \textit{generated} code in isolation, providing a controlled assessment of functional correctness without conflating generation quality with dependency resolution failures. To validate the quality of these test suites, we measured their branch coverage against the human-written ground truth for our 300 evaluation classes, achieving an average of 58\%. This is only marginally lower than Pynguin's reported performance on Python code (63--68\% branch coverage using search-based algorithms~\cite{lukasczyk2022pynguin}), confirming that our test suites provide a robust baseline for assessing functional correctness.

\subsection{Results}

Table~\ref{tab:results} presents results for each LLM across the four quality dimensions.

\begin{table*}[t]
\centering
\caption{Generation quality across four dimensions (300 samples). We report average scores per model.}
\label{tab:results}
\begin{tabular}{lccccc}
\toprule
 & \multicolumn{2}{c}{\textbf{Syntactic}} & \textbf{Semantic} & \textbf{Structural} & \textbf{Functional} \\
\cmidrule(lr){2-3} \cmidrule(lr){4-4} \cmidrule(lr){5-5} \cmidrule(lr){6-6}
\textbf{Model} & ROUGE-L & BLEU & CodeBERTScore-F3 & TSED & Pass Rate \\
\midrule
GPT-o4-mini & 0.60 & 0.46 & 0.88 & 0.82 & 0.29 \\
Claude-4-Sonnet & 0.68 & 0.50 & 0.90 & 0.84 & 0.36 \\
Qwen-3-Coder & 0.67 & 0.49 & 0.90 & 0.84 & 0.35 \\
\midrule
\textbf{Average} & 0.65 & 0.48 & 0.89 & 0.83 & 0.33 \\
\bottomrule
\end{tabular}
\end{table*}

\paragraph{Semantic and Structural Similarity.} Generated classes demonstrate strong semantic similarity (average CodeBERTScore-F3: 0.89) and structural alignment (average TSED: 0.83), indicating that LLMs effectively capture the meaning and architecture of human-written classes. The low standard deviation across models (0.07 for both metrics) suggests consistent high-level understanding regardless of model architecture.

\paragraph{Syntactic Variation.} Syntactic overlap is moderate (average ROUGE-L: 0.65, average BLEU: 0.48), reflecting natural variations in implementation style while preserving semantic intent. LLMs prioritize meaning over exact token reproduction, which is a necessary strength, as multiple valid implementations exist for most class specifications.

\paragraph{Functional Correctness.} Pass rates average 0.33 across the three models, with Claude-4-Sonnet and Qwen-3-Coder achieving 0.35--0.36 and GPT-o4-mini at 0.29. The variance in pass rates demonstrates that OpenClassGen provides meaningful differentiation between models---an essential characteristic for a useful evaluation corpus.

\paragraph{Implications.} These results confirm that class skeletons provide sufficient context for LLM generation. The combination of strong semantic/structural scores with moderate syntactic overlap indicates that LLMs understand class design but vary in surface-level implementation choices. The functional correctness gap (0.33 pass rate vs. 0.89 semantic similarity) highlights the distinction between understanding intent and producing executable code, indicating it as a key challenge for future research.

These results confirm that OpenClassGen enables meaningful differentiation of LLM capabilities across multiple quality dimensions.

\section{Discussion}
\label{sec:discussion}

We discuss how OpenClassGen addresses critical gaps in the current landscape of code generation research, outline concrete use cases, and identify the unique value proposition of our contribution.

\subsection{Addressing the Scale-Realism Trade-off}

Existing class-level datasets force researchers to choose between scale and realism. Synthetic benchmarks like ClassEval~\cite{du2024evaluating} offer controlled evaluation with complete documentation and test suites, but their 100 hand-crafted samples lack the diversity and noise of real-world software. Real-world datasets like RealClassEval~\cite{rahman2025beyond} capture authentic coding patterns, but their 400-sample scale limits statistical power and precludes use for training or fine-tuning.

OpenClassGen resolves this trade-off by providing 324,843 real-world classes, which is three orders of magnitude larger than existing class-level datasets while preserving the documentation variability, structural diversity, and domain heterogeneity characteristic of production software. This scale enables:

\begin{itemize} [leftmargin=*]
    \item \textbf{Statistically robust evaluation}: Large sample sizes support meaningful confidence intervals and effect size estimation across class characteristics.
    \item \textbf{Fine-tuning and training}: Sufficient data volume for supervised fine-tuning of code generation models on class-level tasks.
    \item \textbf{Stratified analysis}: Researchers can subset by documentation level, complexity, or domain while retaining adequate sample sizes per stratum.
\end{itemize}

\subsection{Complementing Evolution-Aware Evaluation}

Recent work has highlighted the importance of temporal considerations in code generation evaluation. HumanEvo~\cite{zheng2024humanevo} demonstrated that evolution-ignored evaluation inflates LLM performance by 10--61\% due to future context leakage and missing deleted code. Their solution rolls back repositories to the commit state when the target functions were written.

OpenClassGen complements this evolution-aware perspective differently. Rather than providing repository context that requires temporal alignment, we extract \textit{self-contained class skeletons} that specify the generation task without external dependencies. This design choice offers two advantages:

\begin{enumerate} [leftmargin=*]
    \item \textbf{Context isolation}: Each class can be evaluated independently without repository rollback infrastructure, enabling scalable evaluation across hundreds of thousands of samples.
    \item \textbf{Controlled prompting}: The skeleton serves as a complete specification, such as what the class should do (docstrings) and how it should be structured (method signatures), without leaking implementation details from surrounding code.
\end{enumerate}

While HumanEvo addresses the question ``Can LLMs generate correct code given realistic repository context?'', OpenClassGen addresses ``Can LLMs generate complete class implementations given only structural specifications?'' These are complementary research questions, and progress on both is necessary for robust LLM-assisted software development.

\subsection{Enabling Structural Analysis}

Unlike existing datasets, OpenClassGen provides 27 static code metrics per class, enabling research directions not possible with prior benchmarks:

\paragraph{Difficulty Prediction.} Our evaluation revealed substantial variance in pass rates, indicating that certain classes are inherently more challenging. The structural metrics enable correlation studies: Which properties predict generation difficulty? Prior work suggests cyclomatic complexity and dependency depth impact LLM performance~\cite{du2024evaluating, rahman2025beyond}, but systematic validation requires large-scale data with per-class metrics, and this is precisely what OpenClassGen provides.

\paragraph{Retrieval-Augmented Generation.} For RAG-based code generation, retrieval quality depends on meaningful similarity measures. Lexical similarity (e.g., BM25~\cite{robertson2009probabilistic}) may retrieve syntactically similar but structurally different classes. OpenClassGen's metrics enable retrieval strategies based on structural properties: retrieve classes with similar complexity profiles, inheritance patterns, or coupling characteristics as few-shot examples.

\paragraph{Curriculum Learning.} Training or fine-tuning LLMs on progressively difficult examples can improve learning efficiency. The structural metrics provide a principled basis for ordering training samples by difficulty, enabling curriculum learning strategies for class-level code generation.

\subsection{Limitations of Existing Benchmarks}

Table~\ref{tab:benchmark_limitations} summarizes limitations of existing benchmarks and how OpenClassGen addresses them. We include class-level datasets (ClassEval, RealClassEval) and repository-level benchmarks that target object-oriented structures (JavaBench, RepoClassBench). Function-level benchmarks such as HumanEvo address orthogonal challenges (temporal evolution) and are discussed in Section~\ref{sec:discussion} but excluded here as their limitations fall outside the scope of class-level generation.

\begin{table*}[t]
\centering
\caption{Limitations of existing class-level benchmarks and how OpenClassGen addresses them.}
\label{tab:benchmark_limitations}
\begin{tabular}{p{0.12\textwidth}p{0.38\textwidth}p{0.42\textwidth}}
\toprule
\textbf{Benchmark} & \textbf{Limitation} & \textbf{How OpenClassGen Addresses It} \\
\midrule
ClassEval & Synthetic data with complete documentation do not reflect real-world variability; 100 samples are insufficient for training or robust evaluation. & 324,843 classes from production repositories with natural documentation variability (19.7\% fully documented, 44.4\% partial, 36.0\% undocumented). \\
\addlinespace
RealClassEval & 400 samples limit statistical power; no structural metrics for difficulty analysis. & Three orders of magnitude larger scale enables robust statistical analysis; 27 static metrics per class support difficulty modelling and stratified evaluation. \\
\addlinespace
JavaBench & Project-level granularity conflates class generation with dependency resolution; Java-only. & Class-level granularity isolates the generation task from dependency management; self-contained skeletons serve as complete specifications. \\
\addlinespace
RepoClassBench & Repository context requirements limit scalability; small sample size. & No repository context required; skeletons are self-contained, enabling evaluation of 324,843 classes without dependency resolution infrastructure. \\
\bottomrule
\end{tabular}
\end{table*}

\subsection{Intended Use Cases}

The following use cases represent intended and anticipated directions enabled by OpenClassGen's scale and metadata. While the current evaluation demonstrates dataset utility for zero-shot generation assessment, experimental validation of fine-tuning, RAG, and other downstream applications is left to future work. Based on our analysis, we identify the following concrete use cases for OpenClassGen:

\paragraph{Use Case 1: Fine-Tuning for Class-Level Generation.} The 324,843 (skeleton, implementation) pairs provide a supervised training signal for future fine-tuning studies. Researchers can fine-tune base models to improve class-level generation, then evaluate on held-out subsets or external benchmarks like ClassEval.

\paragraph{Use Case 2: RAG Corpus Construction.} The structural metadata enables the construction of retrieval corpora where similar classes (by complexity, documentation, or domain) serve as few-shot examples. This is a particularly promising direction for future work, as structural similarity-based retrieval may outperform lexical approaches for class-level generation.

\paragraph{Use Case 3: Failure Mode Analysis.} The scale supports statistically robust analysis of LLM failure patterns. Researchers can correlate structural properties with error types (e.g., do high-coupling classes produce more \texttt{AttributeError}s?) to identify systematic LLM weaknesses.

\paragraph{Use Case 4: Documentation Impact Studies.} The documentation stratification (20\% fully documented, 44\% partial, 36\% undocumented) enables controlled studies of how documentation quality affects generation correctness. Prior work has explored this question through ablation studies that randomly remove docstrings from fully documented classes~\cite{rahman2025beyond}; OpenClassGen instead provides naturally occurring documentation variation, enabling more realistic evaluation. This is a question with direct practical implications for developer workflows.

\paragraph{Use Case 5: Benchmark Subset Curation.} Researchers can use the structural metrics to curate focused benchmarks: e.g., ``high-complexity classes with full documentation'' or ``simple utility classes without docstrings.'' This enables targeted evaluation of specific LLM capabilities.

\section{Limitations}
\label{sec:limitations}

We discuss the limitations of OpenClassGen and suggest possible mitigation strategies.

\paragraph{Data Contamination Risk.} LLMs are trained on public GitHub repositories, and our source projects may overlap with training data, potentially inflating performance estimates. Importantly, the source repositories were collected prior to the training cutoffs of all three evaluated models (GPT-o4-mini, Claude-4-Sonnet, and Qwen-3-Coder), meaning that data leakage cannot be ruled out for any of them. However, existing work~\cite{rahman2025beyond} found no significant performance difference between seen and unseen real-world classes, suggesting memorization plays a minimal role in class-level generation where structural understanding matters more than verbatim recall. We therefore treat the reported results as potentially optimistic upper bounds, and recommend that future evaluations cross-reference repository names and commit timestamps against model-specific training cutoff dates to identify and exclude potentially memorized samples.

\paragraph{Functional Validation Scope.} Our evaluation uses Pynguin for automated test generation, achieving 58\% branch coverage on our evaluation set, which is close to Pynguin's reported performance (63--68\%) on Python code~\cite{lukasczyk2022pynguin}. The remaining 42\% of unexercised branches means that pass rates may underestimate true functional correctness, particularly for stateful or dependency-heavy classes. Additionally, relying on a single automated test generation tool introduces oracle risk: test inadequacy may be conflated with model failure, causing correct implementations to appear incorrect when tests fail to exercise the relevant code paths. Taken together, the reported pass rate of 0.33 should be interpreted as a conservative lower bound rather than a precise measure of functional correctness. Researchers requiring higher-fidelity correctness evaluation can use OpenClassGen's structural metrics to identify classes with simpler dependency structures, or supplement Pynguin with manual test curation for focused studies.

\paragraph{Python-Only Scope.} OpenClassGen contains only Python classes, limiting generalizability to other object-oriented languages such as Java, C\#, or TypeScript. This focus was intentional: Python dominates existing code generation research~\cite{cao2024javabench}, enabling direct comparison with prior benchmarks. That said, the curation pipeline is language-agnostic at the project selection and filtering stages; extending to other languages requires only language-specific AST parsing, which we leave to future work.

Despite these limitations, OpenClassGen's scale, structural metadata, and real-world fidelity provide a robust foundation for advancing class-level code generation research, transforming known constraints into structured opportunities for future investigation.

\section{Conclusion}
\label{sec:conclusion}

We introduced OpenClassGen, a large-scale corpus of 324,843 real-world Python classes extracted from 2,970 engineered open-source projects. Each entry pairs a human-written class implementation with its corresponding skeleton---comprising class and method signatures with associated docstrings---and is complemented with 27 static code metrics.

Our evaluation across three contemporary LLMs (GPT-o4-mini, Claude-4-Sonnet, Qwen-3-Coder) demonstrates that OpenClassGen enables meaningful assessment of class-level code generation. Generated classes exhibit strong semantic similarity (CodeBERTScore-F3: 0.89) and structural alignment (TSED: 0.83) compared to their human-written counterparts, yet functional correctness remains challenging (pass rate: 0.33), aligning with prior findings on real-world class-level generation~\cite{rahman2025beyond}. This gap between similarity metrics and executable correctness confirms the dataset's utility for differentiating model capabilities and studying factors that influence generation quality.

OpenClassGen addresses a critical gap in the code generation landscape: existing class-level datasets are either synthetic and small (ClassEval: 100 classes) or real-world but limited in scale (RealClassEval: 400 classes). By providing three orders of magnitude more samples while preserving real-world documentation variability, structural diversity, and domain heterogeneity, OpenClassGen enables research directions previously infeasible, including fine-tuning for class-level generation, retrieval-augmented generation with structural metadata, and statistically robust failure mode analysis. We hope this corpus serves as a foundation for advancing LLM-assisted code generation toward the structural complexity of real-world software development.

\section*{Data Availability}
\label{sec:replication}
Our replication package, which includes the analysis scripts, can be found here: \url{https://zenodo.org/records/18409150}. The dataset is also available on Hugging Face~\cite{musfiqur_rahman_2025} and can be loaded using the {\tt datasets} Python library as follows:

\begin{lstlisting}[style=pythonstyle, caption={Loading OpenClassGen from Hugging Face.}, label={lst:hf}]
# pip install datasets

from datasets import load_dataset

dataset = load_dataset("mrahman2025/OpenClassGen")
\end{lstlisting}

\begin{acks}
We used AI-powered tools (Claude and Grammarly) to assist with manuscript revisions. All AI-assisted content was reviewed and verified for accuracy by the first author.
\end{acks}

\bibliographystyle{ACM-Reference-Format}
\bibliography{bibliography}

\end{document}